\documentclass[useAMS,usenatbib,rotating]{mnras}
\usepackage{amssymb}     \usepackage{amsmath}    \usepackage{amsfonts}
\usepackage{xspace}    \usepackage{longtable}    \usepackage{graphicx}
\usepackage{times}

\newcommand{\etal}{{\it et~al.\/}}
\newcommand{\HI}{\mbox{\sc Hi}}
\newcommand{\HII}{\mbox{\sc Hii}}
\newcommand{\HIPASS}{{\sc HiPASS}}
\newcommand{\Hline}[1]{\mbox{H{\footnotesize {#1}}}}
\newcommand{\Halpha}{\Hline{\mbox{$\alpha$}}}
\newcommand{\kms}{\mbox{km\thinspace s$^{-1}$}}

\newcommand{\Msun}{\mbox{${\cal M}_\odot$}}

\newcommand{\rhii}{\mbox{$R_{\rm HII}$}}
\newcommand{\rmax}{\mbox{$R_{\rm max}$}}
\newcommand{\siglgr}{\mbox{$\sigma_{{\rm log(}R{\rm )}}$}}
\newcommand{\siglgv}{\mbox{$\sigma_{{\rm log(}V{\rm )}}$}}
\newcommand{\torb}{\mbox{$t_{\rm orb}$}}
\newcommand{\vmax}{\mbox{$V_{\rm max}$}}

\def\lapeq{\mathrel{\hbox{\rlap{\hbox{\lower4pt\hbox{$\sim$}}}\hbox{$<$}}}}
\def\gapeq{\mathrel{\hbox{\rlap{\hbox{\lower4pt\hbox{$\sim$}}}\hbox{$>$}}}}

\title[Cosmic Clocks]{Cosmic clocks: A Tight Radius - Velocity
  Relationship for HI-Selected Galaxies}
\author[G.R.\ Meurer et al.]{Gerhardt R.\ Meurer$^{1}$,\thanks{E-mail:
    gerhardt.meurer@icrar.org} Danail Obreschkow$^{1}$,
  O.\ Ivy Wong$^{1,2}$, Zheng Zheng$^{3}$, 
  \newauthor Fiona M.\ Audcent-Ross$^{1}$, and D.J.\ Hanish$^{4}$\\
$^{1}$International Centre for Radio Astronomy Research, The University of
Western Australia, 35 Stirling Highway,\\ Crawley, WA 6009, Australia\\
$^{2}$ ARC Centre of Excellence for All-sky Astrophysics (CAASTRO), Australia\\
$^{3}$ National Astronomical Observatories, Chinese Academy of Sciences, A20 Datun Road, Chaoyang District, Beijing 100012, China\\
$^{4}$ Spitzer Science Center, California Institute of Technology, MC 220-6, 1200 E California Blvd., Pasadena, CA 91125, USA} 
\begin{document}

\date{Accepted 2018 January 29. Received 2018 January 14; in original form 2017 May 22}

\pagerange{\pageref{firstpage}--\pageref{lastpage}} \pubyear{2018}

\maketitle

\label{firstpage}

\begin{abstract}
{\HI}-Selected galaxies obey a linear relationship between their maximum
detected radius \rmax\ and rotational velocity. This result covers measurements
in the optical, ultraviolet, and \HI\ emission in galaxies spanning a factor of
30 in size and velocity, from small dwarf irregulars to the largest
spirals. Hence, galaxies behave as clocks, rotating once a Gyr at the very
outskirts of their discs. Observations of a large optically-selected sample are
consistent, implying this
relationship is generic to disc galaxies in the low redshift Universe. A linear
$RV$ relationship is expected from simple models of galaxy formation and
evolution. The total mass within \rmax\ has collapsed by a factor of 37 compared
to the present mean density of the Universe. Adopting standard
assumptions we find a mean halo spin parameter $\lambda$ in the range 0.020 to
0.035.  The dispersion in $\lambda$, 0.16 dex, is smaller
than expected from simulations.  This may be due to the biases in our selection of 
disc galaxies rather than all halos.  The estimated mass densities of stars and 
atomic gas at \rmax\ are similar ($\sim 0.5\, M_\odot\, {\rm pc^{-2}}$) indicating 
outer discs are highly evolved. The gas consumption and stellar population build time-scales 
are hundreds of Gyr, hence star formation is not driving the current evolution
of outer discs. 
The estimated ratio between \rmax\ and disc scale length is consistent with long-standing predictions
from monolithic collapse models. Hence, it remains unclear whether disc extent
results from continual accretion, a rapid initial collapse, secular evolution or
a combination
thereof.
\end{abstract}


\section{Introduction}

Based on the Cold Dark Matter (CDM) scenario for galaxy evolution the
main structural and dynamical properties of galaxies' halos and discs are
expected to obey simple virial scaling relations
\citep{fe80,mmw98,dutton+07}. These properties are typically specified
as a radius $R$, rotation velocity amplitude $V$ and a mass $M$, or
alternatively luminosity $L$ as a proxy for mass. For halos in virial
equilibrium we expect $V \propto R \propto M^{1/3}$ \citep[][hereafter MMW98]{mmw98}. While the dark
matter is not directly observable, scaling relations are observed in the
properties of the {\em baryons\/}, although the slopes (power law
exponents) of the relations are not exactly as predicted for the halos
\citep[e.g.][]{courteau+07}.  

The most used scaling relation is the velocity-luminosity relation, better known as the Tully-Fisher Relation \citep[hereafter TFR;][]{tf77}, and similarly the Baryonic Tully-Fisher Relationship \citep{msbd00} which is a velocity-mass relationship. Baryonic physics is messy. The scaling between luminosity and baryonic mass depends on the star formation history which varies between galaxies \citep{grebel97,tolstoy+09,weisz+11a,williams+11}, the Initial Mass Function (IMF) which also apparently varies between galaxies whether they are dominated by young stellar populations \citep{hg08,meurer+09,lee+09hafuv,gunawardhana+11} or old ones \citep{treu+10,dc12,cd12,cappellari+12,smith+12,dms12}, and the dust content and distribution \citep{calzetti+94,gordon+01,tuffs+04}. Theory and observations indicate that feedback from star formation \citep{governato+10,oh+11} or active galactic nuclei \citep[e.g.\ ][]{boneli+16} can rearrange the distribution of baryons, and in the process drag along the dark matter (DM) into an altered distribution, affecting all scaling relations.

The radius-velocity ($RV$) relationship has received somewhat less attention. \citet{courteau+07} and \citet{dutton+07} fit scaling relations to $R$, $V$, and $L$ in a sample of luminous spiral galaxies having optical spectroscopic observations. They found that the scatter in the $RV$ relationship was the highest compared to the $LV$ and $RL$ relationships. Some of the scatter in the $RV$ relation is due to the uncertainties and ambiguities of measuring $R$. This includes lack of a uniform definition of radial scale length \citep[cf.][]{pt06}, contamination by the bulge component, selection effects (especially with surface brightness), and errors due to dust. However, if instead of using a scale length to characterise $R$ we consider an outer radius, then some of these concerns (e.g.\ bulge and dust) are minimised and a tighter relationship can be found. This will allow a better measurement of the intrinsic scatter in the $RV$ relationship which is very sensitive to the spin of the halos in which galaxies lie \citep[e.g.][]{mmw98,dutton+07,courteau+07,og14}.

Here we demonstrate a nearly linear $RV$ relationship in various measurements of {\HI}-selected galaxy samples.  In Section~\ref{s:samp} we present our primary samples and detail the measurements we use. Section~\ref{s:results} shows the observed correlations and quantifies the slopes and scatters; we also test the results on a large comparison sample selected and measured in the optical giving consistent results. In Section~\ref{s:theory} we show what a linear $RV$ relation means in the context of CDM dominated galaxy evolution models.  Our results are discussed further in Section~\ref{s:disc} where we estimate the spin parameter of galaxies, the properties of discs near their outer extents, and then discuss how these results relate to ideas on what limits the extent of galaxy discs. Our conclusions are presented in Section~\ref{s:conc}.

\section{Samples and measurements}\label{s:samp}

\begin{figure*}
\includegraphics[width=\textwidth]{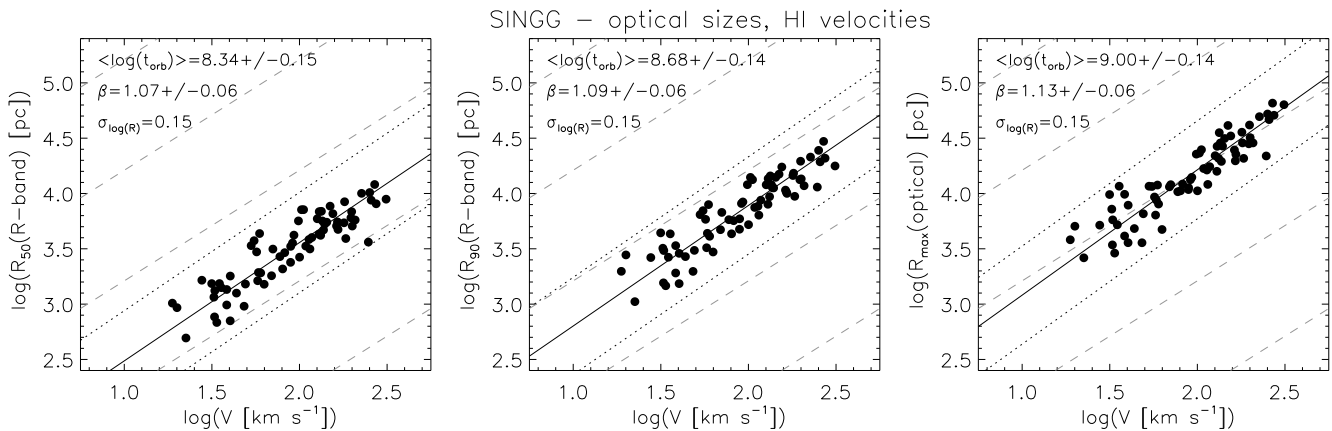}
\caption{Radius $R$ plotted against circular velocity $V$ on a
  logarithmic scale as derived from SINGG optical data. The radii
  plotted in the left-hand, middle and right-hand panels are $R_{\rm 50}({\rm
    R})$, $R_{\rm 90}({\rm R})$, and $\rmax({\rm optical})$,
  respectively, while all three panels plot the same \HIPASS\ \HI\
  based $V$. The solid line shows the iteratively clipped ordinary
  least squares bisector fit to the plotted quantities, while the
  dotted line shows the fit offset by $\pm 3\siglgr$, where \siglgr\
  is the dispersion of the $R$ residuals Each panel is annotated with
  the the mean log orbital time, $\langle\log(\torb)\rangle$ and its dispersion (after
  clipping), the fitted
  slope $\beta$ and its error, and the rms of the residuals in the
  ordinate. The parallel grey dashed lines from bottom to top show
  where $\torb = 10^{8},\, 10^{8.5},\, 10^{9.0},\, 10^{9.5},\,
  10^{10}$ years respectively. \label{f:rvop}}
\end{figure*}

 We measure the $RV$ relationship in three primary samples.  The first uses optical data from the Survey of Ionization in Neutral Gas Galaxies \citep[SINGG;][]{meurer+06}, which is an \Halpha\ and R-band follow-up survey to the \HI\ Parkes All Sky Survey \citep[\HIPASS;][]{meyer+04,zwaan+04,koribalski+04} combined with single dish \HI\ data from \HIPASS. The second uses data from the Survey of Ultraviolet emission in Neutral Gas Galaxies \citep[SUNGG;][]{wong07}, which observed {\HIPASS}-selected galaxies in the ultraviolet (UV) with GALEX, for a sample largely overlapping with SINGG. Here, we use sub-samples of SINGG and SUNGG designed to ensure that reasonable rotation amplitudes can be derived from the \HIPASS\ \HI\ data.  Specifically, both samples are selected to have major to minor axial ratios $a/b \geq 2$ and to be the only apparent star forming galaxy in the system. The $a/b$ cut guarantees a minimum inclination of about 60$^\circ$, thus limiting projection errors in calculating orbital velocities. For those galaxies observed by SINGG, the isolation criterion was determined using the \Halpha\ images, which are roughly the same size as the \HIPASS\ beam. For those SUNGG galaxies not observed by SINGG, isolation was determined morphologically; systems with companions of similar angular size, obvious signs of interaction, or noted as interacting with another galaxy in the literature were excluded. These selection criteria results in 71 and 87 galaxies from SINGG and SUNGG respectively, with an overlap of 47 galaxies in common.  The third sample uses \HI\ imaging data of the 20 galaxies studied by \citet[][hereafter MZD13]{mzd13}.  

In all three samples, \HI\ data is used to infer the maximum rotation amplitude.
The implicit assumption is that the \HI\ in these galaxies is dominated by a
rotating disc.  It is important to bear in mind that the selection of the samples requires detectable amounts of \HI, and thus is biased against gas-poor disc galaxies
(e.g.\ S0 galaxies and ellipticals).  Note that the $a/b$ cut applied to the
SINGG and SUNGG samples also is likely to remove early type and S0 galaxies from
our samples. As pointed out by \citet{meurer+06}, very few such galaxies are
found in the SINGG sample. The selection against early type
galaxies may have implications on the types of halos they are
associated with, as discussed in \S\ref{ss:spin}.  As we show below, the implied rotational amplitudes range from $\sim 10$ \kms\ (dwarf galaxies) to $\sim$300 \kms\ (the largest spirals).

The radii used for the SINGG and SUNGG samples depend on the {\em maximum\/} extent of the galaxies observed in the optical and UV, respectively. Both surveys are designed to measure the total light of extended nearby galaxies using a series of concentric elliptical annuli.  For SINGG the apertures are set in a manner slightly modified from that given in \citet{meurer+06}.  As noted there, the aperture shape ($a/b$ and position angle) and centre are set by eye to include all the apparent optical emission.  In most cases, this shape matches well the apparent shape of the galaxy in the R-band, i.e.\ a tilted disc.  We then grow the apertures to an arbitrarily large size, and determine, by eye, where the raw ({\em before\/} sky subtraction) radial surface brightness profile levels off. The surface brightness of the galaxy at that radius is on the order of 1\%\ of the sky brightness.  

The radius where the raw surface brightness profiles flatten is called the
maximum radius \rmax. Since the R-band light almost always can be traced
further than \Halpha, \rmax\ typically measures the maximum detectable
extent in the optical continuum. Most exceptions are dwarf galaxies with strong
minor-axis outflows. Optical sizes were estimated in this manner by two of us.
First by DH and then by GRM who ``tweaked'' the size estimates in about half of
the SINGG sample. Typically those that were adjusted were made larger because
the raw profiles indicated a small amount of additional flux
could be gained doing so. Here we use the tweaked aperture radii. Compared to
using the initial estimates, the use of the tweaked apertures increases \rmax\
by 0.06 dex on average and also reduces the scatter in the residuals of the fits
described below by 0.06 dex (when taken in quadrature). The SUNGG maximum radius
is set in a similar manner; it is determined separately in NUV and FUV and the
maximum of the two is taken as \rmax.

For both the SINGG and SUNGG samples we
interpolate enclosed flux versus aperture semi-major axis profiles to
determine the radii containing 50\%\ ($R_{\rm 50}$) and 90\%\ ($R_{\rm
  90}$) of the flux in the R-band and UV, respectively.

For the MZD13 sample we use three radii: \rmax\ is the maximum extent of
the \HI\ radial profiles as given in the original studies used by MZD13,
while $R_1$ and $R_2$ represent the extent of the region of the \HI\
surface mass density profile $\Sigma_{\rm HI}$ fit with a power law by
MZD13. These radii are set by eye to mark kinks in the \HI\ radial profiles, indicating changes of slope in $\log(\Sigma_{\rm HI}(R))$.  On average they are close to the radii that contain 25\%\ and 75\%\ of the \HI\ flux respectively (MZD13).  Unfortunately, neither MZD13, nor the studies they employed, calculated $R_{\rm 50}$ and $R_{\rm 90}$ for the \HI\ data.

The shape of the Rotation Curve (hereafter RC) $V(R)$ of galaxies varies systematically with mass, or peak rotational velocity, from nearly solid body (linearly rising) for the lowest mass galaxies, to RCs that are flat at nearly all radii, or even slightly declining at large $R$ for the most massive galaxies \citep{ps91,pss96,cgh06}. Unless stated otherwise we take $V$ to be the maximum rotational amplitude.  For most cases this will be the amplitude at the flat part of the RC.  In the majority of other cases it will be the farthest measured point of the RC.  We take these definitions to be synonymous. For the SINGG and SUNGG samples we derive $V$ from the full width at half maximum of the \HI\ spectrum from \HIPASS, assuming a flat RC over all relevant radii.  We follow the method of \citet{meyer+08} and correct the line widths for inclination, and broadening resulting from turbulence, relativity, instrumental effects and data smoothing. As with \citet{meyer+08} the inclinations are derived from $a/b$.  For the \HI\ sample we interpolate the RCs, from the various original studies used by MZD13 to arrive at rotation amplitudes at $R_1$, $R_2$, and \rmax\ separately (i.e.\ $V(R_1)$, $V(R_2)$, and $V(R_{\rm max})$).

\section{Results}\label{s:results}

\subsection{Observed Correlations}\label{ss:obscor}

\begin{figure*}
\includegraphics[width=\textwidth]{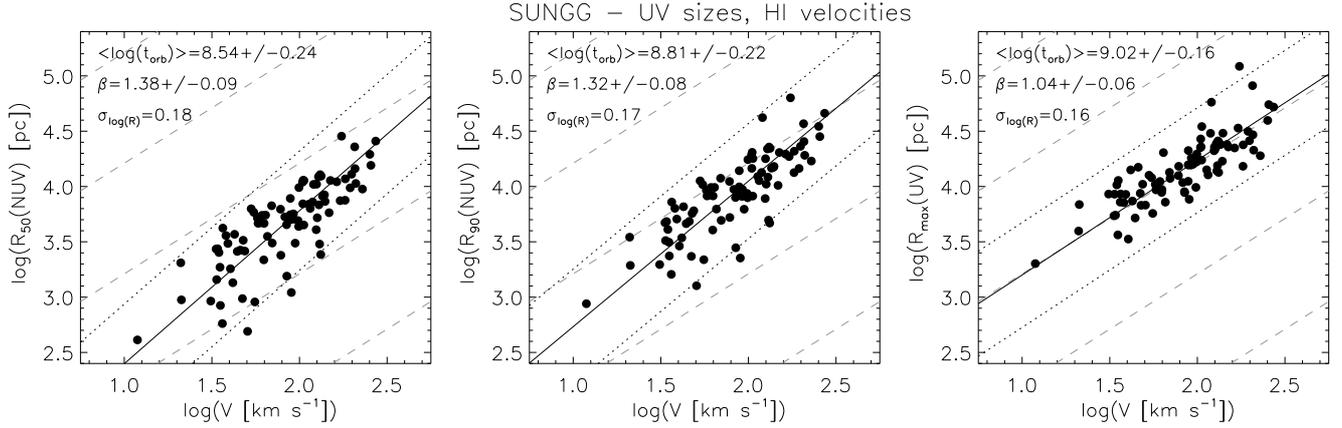}
\caption{Radius $R$ plotted against circular velocity $V$ on a
  logarithmic scale as derived from SUNGG ultraviolet data. The radii
  used here are $R_{\rm 50}({\rm NUV})$, $R_{\rm 90}({\rm NUV})$, and
  $\rmax({\rm UV})$ in the left-hand, middle and right-hand panels,
  respectively. The meanings of the various lines and annotations are
  the same as in Fig.~\ref{f:rvop}.
  \label{f:rvuv}}
\end{figure*}

\begin{figure*}
\includegraphics[width=\textwidth]{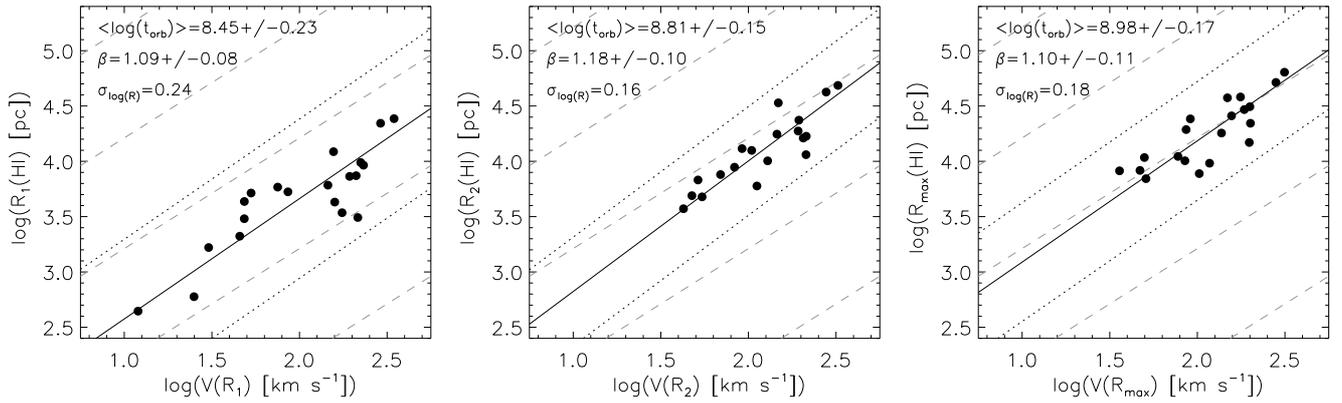}
\caption{Radius $R$ plotted against circular velocity $V$ on a
  logarithmic scale for the \HI\ sample of MZD13.  The radii $R_1$ (left-hand panel)
  and $R_2$ (centre panel) delimit the region where the \HI\ surface brightness
  profile is a power-law with index $\gamma \approx -1$ (see MZD13 for
  details), while $R_{\rm max}$ (right-hand panel) is the maximum detected
  extent of \HI. The meanings of the various lines and annotations are
  the same as in Fig.~\ref{f:rvop}.
  \label{f:rvhi}}
\end{figure*}

We show the observed correlations separately for each data set in three figures. Figure~\ref{f:rvop}
shows the $RV$ relationship for the SINGG optical data.  In the left
panel the $y$-axis gives the radius as $R_{\rm 50}({\rm R})$, i.e. the radius
containing 50\%\ of the R-band light; similarly the middle panel
shows $R_{\rm 90}({\rm R})$ as the radius; while the right panel uses \rmax\
as defined from the SINGG optical data. The velocity in all panels is
the circular velocity $V$ defined from the \HIPASS\ \HI\ line widths
(\S\ref{s:samp}).  Similarly, Fig.~\ref{f:rvuv} shows
$R_{\rm 50}({\rm NUV})$, $R_{\rm 90}({\rm NUV})$, and \rmax\ from
the SUNGG UV data in the left, middle and right panels, respectively,
against $V$ derived from \HIPASS.  Figure~\ref{f:rvhi} shows
the \HI\ radii $R_1$, $R_2$ and \rmax\ plotted against the
circular velocities interpolated at those radii $V(R_1)$,
$V(R_2)$, $V(\rmax)$ in the left, middle and right panels
respectively.

We fit the $RV$ relations in log-log space as
\begin{equation}
\log(R) = \alpha + \beta \log(V)
\end{equation}
using an ordinary linear least squares bisector algorithm \citep{isobe+90} weighting each point equally, and iteratively clipping points that deviate from the fit by more than three times the dispersion in $R$. Table~\ref{t:fitp} reports the results of the fits, giving the coefficients $\alpha$, $\beta$, the dispersion of the residuals \siglgr, \siglgv, and Pearson's correlation coefficient $r_{xy}$. Some of these quantities are also listed in Figures~\ref{f:rvop} - \ref{f:rvhi}.

\begin{table*}
\caption{Fit parameters}\label{t:fitp}
\begin{center}
\begin{tabular}{lccccccccc}
\hline \hline
Sample & radius & $N_{\rm used}$ & $N_{\rm rej}$ & $\alpha$ & $\beta$    & $\log(t_{\rm orb})$ & \siglgr & \siglgv & $r_{\rm xy}$ \\
\multicolumn{1}{c}{(1)} & (2) & (3) & (4) & (5) & (6) & (7) & (8) & (9) & (10) \\
\hline 
SINGG   & $R_{\rm 50}({\rm R})$  & 71 &  0 & $1.42\pm 0.12$ & $1.07\pm 0.06$ & $8.34\pm 0.15$ & 0.15 & 0.14 & 0.899 \\
SINGG   & $R_{\rm 90}({\rm R})$  & 71 &  0 & $1.71\pm 0.12$ & $1.09\pm 0.06$ & $8.68\pm 0.14$ & 0.15 & 0.14 & 0.908 \\
SINGG   & $R_{\rm max}({\rm opt})$ & 71 &  0 & $1.95\pm 0.12$ & $1.13\pm 0.06$ & $9.00\pm 0.14$ & 0.15 & 0.13 & 0.914 \\ 
SUNGG   & $R_{\rm 50}({\rm NUV})$  & 88 &  0 & $1.11\pm 0.11$ & $1.33\pm 0.08$ & $8.54\pm 0.22$ & 0.18 & 0.24 & 0.797 \\ 
SUNGG   & $R_{\rm 90}({\rm NUV})$  & 88 &  0 & $1.56\pm 0.15$ & $1.24\pm 0.08$ & $8.81\pm 0.21$ & 0.18 & 0.22 & 0.805 \\ 
SUNGG   & $R_{\rm max}({\rm UV})$ & 87 &  1 & $2.16\pm 0.12$ & $1.04\pm 0.06$ & $9.02\pm 0.16$ & 0.16 & 0.16 & 0.830 \\
\HI\      & $R_{\rm 1}({\rm HI})$  & 20 &  0 & $1.49\pm 0.15$ & $1.09\pm 0.08$ & $8.45\pm 0.23$ & 0.24 & 0.22 & 0.859 \\
\HI\      & $R_{\rm 2}({\rm HI})$  & 19 &  1 & $1.64\pm 0.19$ & $1.18\pm 0.10$ & $8.81\pm 0.15$ & 0.16 & 0.14 & 0.876 \\
\HI\      & $R_{\rm max}({\rm HI})$ & 20 &  0 & $1.99\pm 0.24$ & $1.10\pm 0.11$ & $8.98\pm 0.17$ & 0.18 & 0.16 & 0.825 \\
PS1     & $R_{\rm 50}({\rm r})$  & 692 &  6 & $1.32\pm 0.07$ & $1.07\pm 0.03$ & $8.26\pm 0.13$ & 0.14 & 0.13 & 0.770 \\
PS1     & $R_{\rm b}({\rm r})$  & 694 &  4 & $1.41\pm 0.07$ & $1.14\pm 0.03$ & $8.52\pm 0.14$ & 0.15 & 0.13 & 0.761 \\
PS1     & $R_{\rm 90}({\rm r})$  & 689 &  9 & $1.68\pm 0.06$ & $1.07\pm 0.03$ & $8.62\pm 0.12$ & 0.12 & 0.11 & 0.818 \\ 
\hline 
\end{tabular}
\end{center}
Column (1): the Galaxy sample fitted.
Column (2): the radius measured.
Column (3): the number of data points used in the fit.
Column (4): the number of data points rejected from the fit. 
Column (5): the zeropoint of the fit.
Column (6): the slope of the fit.
Column (7): average log of the orbital time of the fitted data points.
Column (8): dispersion in the log of the residuals in radius $R$ of the fitted points. 
Column (9): dispersion in the log of the residuals in orbital velocity $V$ of the fitted
points (or implied orbital velocity $V'$ for the PS1 sample).
Column (10): Pearson's correlation coefficient using all data points.
\end{table*}

For the optical and UV samples the fits are the ``best'' at \rmax, where best is defined as having the highest $r_{xy}$ and lowest \siglgr\ and \siglgv. For the \HI\ sample, the fit at $R_1$, marking where the \HI\ profiles flatten, is much worse than the other two fits.  The flattening is likely to be due to the increasing dominance of molecular gas at small radii \citep{leroy+08,bigiel+08}. The fits at $R_2$ and \rmax\ have similar scatters, indistinguishable statistically.  In summary, the fits are their best, or close to it, at \rmax\ where $\beta$ is close to but slightly greater than unity, that is, a linear relationship.

\begin{figure*}
\includegraphics[width=\textwidth]{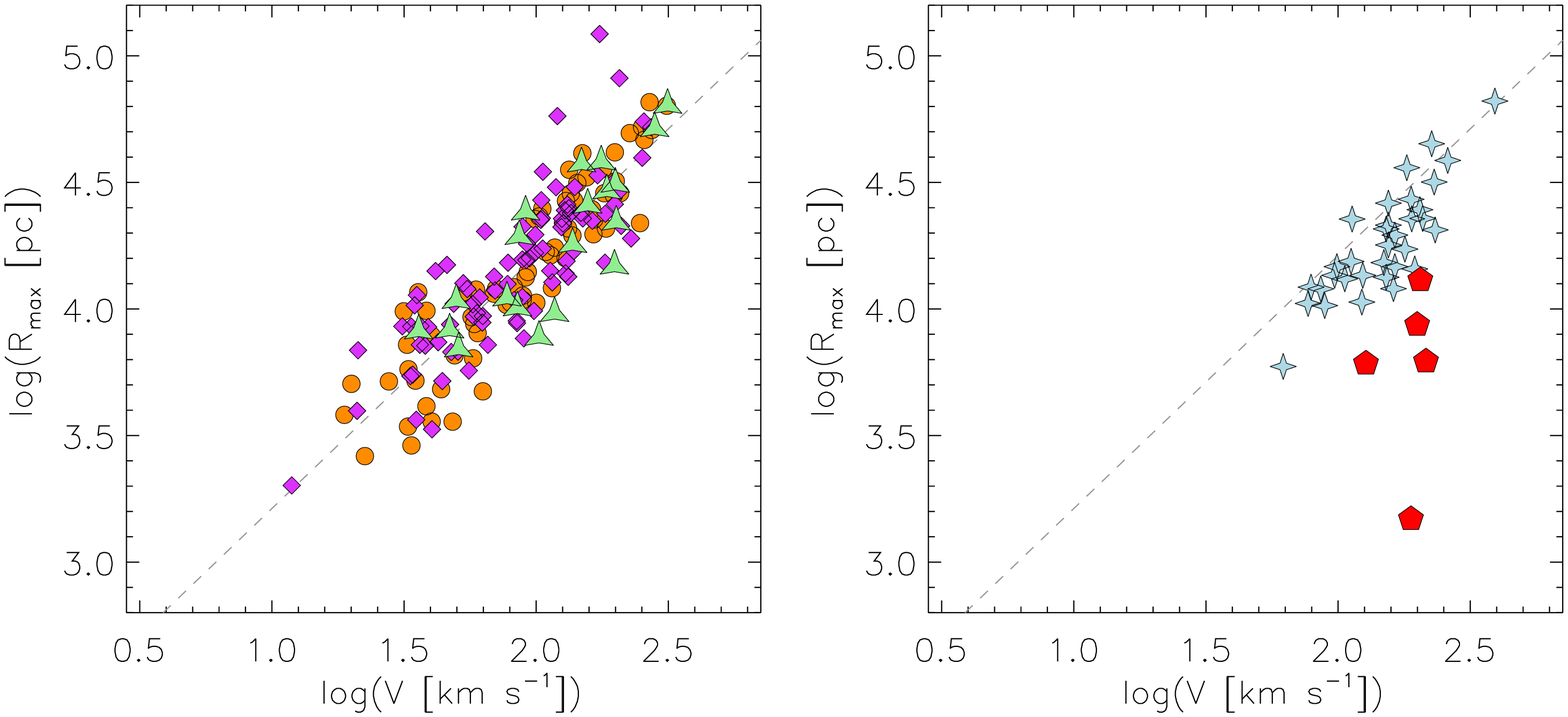}
\caption{Maximum radius \rmax\ plotted against circular velocity $V$ on a logarithmic scale for our combined primary samples is shown in the left panel. Here orange circles, purple diamonds, and green triangles show the SINGG, SUNGG, and \HI\ samples displayed in Figures~\ref{f:rvop}c, \ref{f:rvuv}c, and \ref{f:rvhi}c respectively. In the right panel we show other estimates of the maximum disc size compared to rotational amplitude. In the right panel, the blue four pointed stars shows the optically determined \rmax\ of edge-on galaxies for the sample of \citet{kkg02} against their $V$. The red pentagons show the \HII\ region truncation radius \rhii\ plotted against $V$ for galaxies in common between the samples of \citet{mk01} and MZD13. The grey dashed line in both panels shows the relationship expected for $\torb = 1$ Gyr.  
  \label{f:rvall}}
\end{figure*}

 A linear $RV$ implies that the orbital time
\begin{equation}
\torb = \frac{2\pi R}{V} \label{e:rhotorb}
\end{equation}
is constant (assuming the orbit shape is well approximated by a circle). We list the mean $\log(\torb)$ in Table~\ref{t:fitp} and
the panels of Figures~\ref{f:rvop} - \ref{f:rvhi}. The $RV$ relation
at \rmax\ is nearly identical in the three figures even though
\rmax\ is defined at very different wavelengths, which are sensitive to
different physical processes. Figure~\ref{f:rvall}a over-plots the
three samples at \rmax, showing the excellent correspondence in the $RV$
relationships.  They all imply that $\torb \approx 1$ Gyr, with a
scatter of 0.14 to 0.18 dex (38\%\ to 51\%). Thus, {\HI}-selected disc
galaxies behave like clocks and rotate once in a Gyr at their
outermost detected radii, for galaxies which range in radius from
$\rmax \sim 1.5$ kpc, having $V \sim 10\, km\, s^{-1}$ to galaxies
with $\rmax \sim 50$ kpc and $V \sim 300\, km\, s^{-1}$.  The $RV$
relationship for $\torb = 1$ Gyr is shown with the dashed line in
Figure~\ref{f:rvall}a.

The SINGG--$RV$ relation is equally well defined at $R_{50}$, $R_{90}$ and \rmax.  However, the meaning is less clear when using $R_{50}$ and $R_{90}$. The velocity used, $V$, is determined from the line width of integrated \HI\ velocity profiles of galaxies that are spatially unresolved. The \HI\ in galaxies typically is weighted to larger radius than the easily observed optical emission \citep[e.g.\ ][]{leroy+08}, hence the derived $V$ is also applicable to large radii.  The rotation amplitude at $R_{90}$ and $R_{50}$ will be systematically over-estimated using $V$ as one goes to lower rotation amplitudes and shorter radii (i.e.\ the effect will be stronger for $R_{50}$ than $R_{90}$).  Hence, if we used the true $V$ values at $R_{90}$ and $R_{50}$ then we should see shallower $\beta$ values than shown in Fig.~\ref{f:rvop}.

The $RV$ relations in the UV also are defined using 
\HI\ velocity profiles.  Here we see significantly larger $\sigma_{\rm
  log(R)}$ residuals when using $R_{50}$ and $R_{90}$ as well as
steeper $\beta$ values compared to the $RV$ relation at \rmax. We
posit that the worse fits are due to whether or not galaxies have a
central starburst, and the degree to which they are affected by dust.
These will have more of an impact on the distribution of the UV
luminosity at small radii than in the determination of \rmax.

\subsection{Results for a comparison sample}\label{ss:compsamp}

Our primary results are for three samples having {\HI}-based $V$ measurements, one has {\HI}-based $R$ measurements, and two have overlapping selections based on \HI\ properties. In order to address whether our results may be a byproduct of working with \HI\ data, we now consider a sample that is selected and measured in the optical; the sample of 698 disc galaxies of \citet{zheng+15}.  The sample was selected from the Pan-STARRS1 (PS1) Medium Deep Survey \citep{chambers+16} fields, and measured from the stacked survey images. The galaxies were selected to have images in all PS1 bands (g,r,i,z,y), spectroscopic redshifts from SDSS-III\footnote{http://skyserver.sdss.org/}, to be fairly face-on ($a/b < 2$), and to be well resolved with a \citet{petrosian76} radius\footnote{where the local surface brightness is 20\%\ of the average interior surface brightness} $R_P > 5''$.  Galaxy profiles are then measured to $2R_P$. This algorithm recovers $\ge 94\%$ of the total light for galaxies having a \citet{sersic63} index $n \le 2$ typical of disc galaxies \citep{graham+05}.  \citet{zheng+15} found that radial surface brightness profiles typically show a ``break'', or change in slope, in the bluer bands with the break less apparent towards longer wavelengths. They fitted stellar population models to annular photometry in the five bands to derive stellar mass density profiles and integrated to yield the total stellar mass.  They recorded $R_{50}$, $R_{90}$ and the break radius $R_b$ all measured in the r-band.  Hence, as with the other samples, we have three fiducial radii to work with.  Instead of using a measured rotational velocity, we use $V'_c$, the circular velocity estimated from the stellar mass based TFR of \citet{reyes+11}.  This fit to the TFR has been shown to well represent the kinematics of an SDSS based sample \citep{simons+15} which has redshifts similar to this PS1 sample.

\begin{figure*}
\includegraphics[width=175mm]{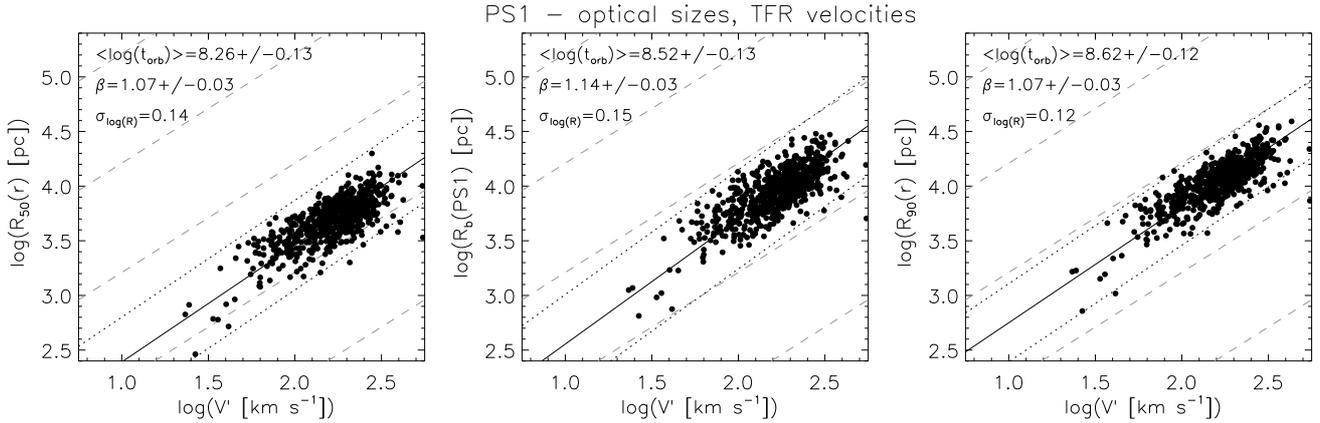}
\caption{Radius $R$ plotted against estimated circular velocity $V'$ on a
  logarithmic scale for the Pan-STARRS1 sample of \citet{zheng+15}.  Here $V'$ is not 
  a direct measurement of circular velocity, but estimated from the stellar mass based TFR 
  of \citet{reyes+11}.  The radii
  used here are the radius containing 50\% of the PS1 r-band light $R_{\rm 50}({\rm r)})$, 
  the break radius $R_b({\rm PS1})$ and the radius containing 90\% of the r-band light $R_{\rm 90}({\rm r})$
  in the left-hand, middle and right-hand panels,
  respectively. The meanings of the various lines and annotations are
  the same as in Fig.~\ref{f:rvop}. 
  \label{f:rvps}}
\end{figure*}

The resulting $RV'$ relationships are shown in Fig.~\ref{f:rvps}.  We fit this sample in the same manner as done for the other samples (\S\ref{ss:obscor}). The fit parameters are tabulated in Table~\ref{t:fitp}.  In all three cases the fits to the PS1 data are nearly linear ($\beta \approx 1$), with the fit at $R_{90}$ being closest to linear and having the smallest scatter $\sigma_{\rm log(R_{\rm 90})} = 0.12$ dex of any of our $RV$ fits.   

Amongst the PS1 sample fits, the one at $R_b$ has the largest scatter, 0.15 dex in $\sigma_{{\rm log\/}(R_b)}$, and deviates the farthest from linear ($\beta = 1.12$).  Nevertheless, the scatter about the mean orbital time of 0.13 dex compares well with the other $RV$ fits.  The larger scatter compared to the fit at $R_{90}$ may arise because the strength of the radial profiles breaks is highly variable with some galaxies ``breaking down'' (more typical), others ``breaking up'', and some showing no discernible breaks \citep{freeman70,pt06,zheng+15}.
 
The (logarithmic) mean \torb\ at $R_{50}$ and $R_{90}$ for the PS1 sample 0.18, 0.42 Gyr respectively, is close to that for the SINGG sample 0.22, 0.48 Gyr.  Meanwhile, at $R_b$ the mean $\torb = 0.36$ Gyr, intermediate between that at $R_{50}$ and $R_{90}$. Hence $R_b \approx 1.8 R_{50} \approx 0.8 R_{90}$ for the PS1 sample. While \citet{zheng+15} do not measure \rmax, they note that typically $R_{90} = R_P$ and that most of the light is recovered at $2R_P$, hence we expect $\rmax \approx 2R_{90}$ for the PS1 sample. For flat RCs, then we infer that $\torb(\rmax) = 2\torb(R_{90}) = 0.83$ Gyr for the PS1 sample, within 0.08 dex of the SINGG sample.  The \torb\ estimates for the PS1 sample at $R_{50}$, $R_{90}$ and that implied at \rmax\ are all lower than those for the SINGG sample, suggesting a more general optical selection of galaxies may result in smaller galaxies than an \HI\ selection.  However, the differences are all close to or about equal to the 0.06 dex systematic error noted in \S\ref{s:samp}. Hence to that level of accuracy, the same $RV$ relationship for \HI\ selected galaxies applies to all disc galaxies at low redshifts.  

The somewhat tighter fits to the PS1 sample does not necessarily mean that the intrinsic scatter in the $RV$ relations is lower than for the SINGG sample. This is because an inferred rather than measured velocity is used. Since the $V'$ in Fig.~\ref{f:rvps} is derived from luminosities, these are essentially $RL$ or $RM_\star$ correlations we are showing. \citet{ss11} find a very tight $RL$ relation (having a scatter of 0.05 dex in $R$) using their SFI++ sample of spiral galaxies, and employing isophotal radii and I-band luminosities.  Similarly, both \citet{courteau+07} and \citet{hall+12} find smaller scatters in their $RL$ relations than their $VR$ relations.  In part, this is because errors in $L$ are effectively reduced by a factor of 3 to 4 due to the TFR scaling, making them competitive or better than velocity errors \citep{ss11}. Velocities are also more prone to errors in inclination, position angle and non-circular motions.  Working with stellar mass (fitted to photometry), as done with our PS1 sample, also improves the accuracy by effectively spreading any error over five bands and weighting the results to the reddest bands.  But improved accuracy may not be the only cause for the tight fits in Fig.~\ref{f:rvps}.   \citet{ss11} performed a careful error analysis of the scatter in their scaling relations and estimate the intrinsic scatter in their $RL$ relations ($\sim 0.034$ dex in $R$) is less than half of that in their $RV$ relations ($\sim 0.084$ dex in $R$).  

\section{Expectations from simple galaxy evolution
  models}\label{s:theory}

A constant \torb\ at \rmax\ implies a constant spherically averaged mean mass (baryons and DM) density
$\rho$ interior to \rmax\ since
\begin{equation}
\rho = \frac{3\pi}{G\torb^2} 
\end{equation}
where $G$ is the gravitational constant.  Our adopted
$\torb(\rmax) = 1$ Gyr yields the mean mass density interior to \rmax: 
\begin{equation}
\langle\rho(\rmax)\rangle = 2.1\times 10^{-3}\, \Msun\, {\rm pc^{-3}}.\label{e:rhormax}
\end{equation}

The closure density of the universe, $\rho_c$, is given by
\begin{equation}
\rho_c = \frac{3H_z^2}{8\pi G} \label{e:rhoc}
\end{equation}
where $H_z$ is the redshift ($z$) dependent Hubble constant. This can be used to estimate the collapse factor of matter within \rmax. Adopting $H_0 = 70\, {\rm km\, s^{-1}\, Mpc^{-1}}$ and the results of the Planck Collaboration \citep{planck16_2014} that the ratio of the cosmic matter density to the closure density $\Omega_M = 0.315$ then we have
\begin{equation}
\frac{\langle\rho(\rmax)\rangle}{\Omega_M\rho_c} = 49000.
\end{equation}
The third root of this is the average collapse factor $f_c(\rmax)$ of the matter
within \rmax\ compared to the present day matter density of the universe:
\begin{equation}
\langle f_c(\rmax)\rangle = 36.6.
\end{equation}

The ``virial radius'' $R_{200}$ is usually defined as the radius where
the mean density of the enclosed mass is 200 times larger than $\rho_c$. From
eq.~\ref{e:rhoc} it is apparent that $\rho_c$ depends only on
redshift, and thus, by definition, a linear $RV$ relationship is expected
at $R_{200}$ at any given epoch given by eq.\ 2 of MMW98:
\begin{equation}
R_{200} = \frac{V}{10 H_z}. \label{e:rvvir}
\end{equation}
From eq.~\ref{e:rhotorb} the orbital time at the virial radius at the
present epoch is $\torb = 8.8$ Gyr.  

The RC interior to $R_{200}$ depends on the distribution of DM and baryons. MMW98 used an analytical approach to examine the expected structure of disc galaxies within DM halos under a variety of plausible assumptions about cosmogeny and distribution of the baryons and DM.  They adopt a simple isothermal sphere to parameterise DM halos, and show that this framework is convenient for understanding how the galaxy scaling relations are influenced by the properties of their halos.  Adopting this approach MMW98 derived eq.~\ref{e:rvvir}, above. An isothermal sphere has a flat (constant) RC and a density profile
\begin{equation}
\rho(R) = \frac{V^2}{4\pi G R^2} . \label{e:rhoiso}
\end{equation}
A flat RC is well supported observationally in most disc
galaxies, especially at large radii \citep[e.g.][]{rft1978,bosma1981b,mfb92,deblok+08,epinat+08}, while
the shape of the inner part of the RC varies systematically
with mass, or $V$ \citep[e.g.][]{ps91,cgh06}.  Since our results are the most consistent at \rmax,
where RCs are typically flat, the isothermal
approximation suffices for our purposes. For a pure exponential disc galaxy in a dominant isothermal halo MMW98 derive the disc scale length
$R_d$ relative to $R_{200}$ in their eq. 12.  Combining that with
eq.~\ref{e:rvvir} yields
\begin{equation}
R_d = \frac{V}{\sqrt{200}H_z}\left(\frac{j_d}{m_d}\right)\lambda \label{e:rdv}
\end{equation}
where $j_d$ is the fraction of the total angular momentum in the disc, $m_d$ is the fraction of the total mass in the disc, and $\lambda$ is the spin parameter. For systems that are not purely exponential discs in an isothermal halo, the scale factor ($1/[\sqrt{200}H_z]$) will vary depending on the detailed distributions of DM and baryons (MMW98). Thus a linear relationship between $R_d$ and $V$ should exist if $(j_d/m_d)\lambda$ is constant. 

If the DM and baryons are well mixed when galaxies collapse one would naively
expect $j_d = m_d$ and thus a constant $j_d/m_d$ (MMW98). This is also the
working assumption of \citet{fe80} whose simple models were consistent with the
observations of the time. While it is impossible to observationally confirm this
expectation because of the invisible nature of DM, simulations allow it to be
tested. \citet{posti+17}, using a similar approach to ours, and matching of
galaxy properties to halo properties from a variety of recent cosmological
N-body simulations, find that $j_d/m_d^{2/3}$ is approximately constant, close
to what what we require. We note that a constant $j_d/m_d$ is difficult to
reproduce in more detailed numerical simulations \citep{governato+10}. The
dependence of $\lambda$ on mass and other parameters for dark halos, as measured
in numerous simulations, is also weak \citep[e.g.
MMW98;][]{cl96,maccio+07,bett+07}. Thus naive considerations tell us that we
expect a linear $RV$ relation when a disc scale length, or anything proportional
to it, is used to measure size. This would be the case for the isophotal sizes
of pure exponential discs that have constant central surface brightness as
originally proposed by \citet{freeman70}.

Disc galaxies, however, are not that simple. They typically contain a bulge increasingly apparent with morphological type \citep[e.g.][]{hubble26}. Since Freeman's landmark work, it has become apparent that discs obey a surface brightness -- luminosity relationship \citep[e.g.][]{kauffmann+03b}, and that the radial profiles frequently show breaks from being pure exponential \citep[][]{freeman70,pt06,zheng+15}. However, allowing for these complications may not necessarily cause major changes to the the $RV$ relation. MMW98 derive the behaviour of an exponential disc in a halo having the typical profile found in CDM-only simulations \citep[][hereafter NFW]{nfw97} which is allowed to respond to the disc's mass.  They find relationships for $R_d$ and the maximum rotational velocity that differ from the isothermal case by form functions that depend on $j_d$, $m_d$, $\lambda$ and halo concentration $c$. Of these, $c$ is the parameter that is expected to have the largest impact \citet{dutton+07}. For example, MMW98 considered the case of a bulge plus disc embedded in an NFW halo, and found disc size depends on assumptions about angular momentum transfer between the bulge, disc and halo. They found disc sizes can vary by a factor of about two, while maximum velocities only vary by $\lapeq 20$\%.

\section{Discussion}\label{s:disc}

The formalism presented in Section \ref{s:theory} allows us to place our results in a
cosmological context. We continue with this approach in Section \ref{ss:spin} by
examining the constraints on the spin parameter $\lambda$ and its dispersion implied by our results. Section
\ref{ss:props} discusses what our results imply for the properties at the disc
outskirts. Section \ref{ss:rmax} argues that our results are best explained by a
true physical truncation of discs. While the formalism presented thus far
implies continual accretion limits the extent of discs, Section \ref{ss:alttrunc}
considers other scenarios for limiting the extent of discs. Finally we present
some ancillary implications of our results in Section \ref{ss:imp}.

\subsection{Spin Parameter}\label{ss:spin}

Equation \ref{e:rdv} is readily re-arranged to be 
\begin{equation}
\lambda = \frac{\sqrt{50}}{\pi}\frac{\torb(R)}{t_{\rm H}}\frac{R_d}{R} \label{e:lambda}
\end{equation}
where $\torb(R)$ is the orbital time at radius $R$, and $t_{\rm H} = H_z^{-1}$
is the Hubble time (13.96 Gyr for our $H_0$), and assuming $j_d/m_d = 1$. Thus,
spin parameter can be estimated from the orbital
time at a given radius and the scaling of that radius with disc scale length.
Since $R_d$ was not measured in our samples, indirect estimates of this scaling
must be made.  We do this using two approaches. 

First, if all baryons are in an un-truncated exponential disc, we can use the SINGG results shown in Fig.~\ref{f:rvop} and Table~\ref{t:fitp} to estimate $\lambda$. Noting that the radius containing 50\%\ and 90\%\ of the light of such a disc corresponds to 1.68 and 3.89 times $R_d$, and converting the mean orbital times in the log from Table~\ref{t:fitp}, then we have $\lambda = 0.021, 0.020$ estimated from $\torb(R_{50})$ and $\torb(R_{90})$ respectively. Being virtually identical, we take $\lambda = 0.020$ to be the average spin under the pure exponential disc assumption.

Second, we estimate $\rmax/R_d$, and thus $\lambda$ by scaling 
from the sample of \citet{kkg02} shown in Fig.~\ref{f:rvall}b. They fit models
including both an exponential disc and bulge to the light distribution of
edge-on galaxies. Their disc model is truncated, yielding a maximum radius
\rmax, which they find to be on average a factor $3.6\pm 0.6$ times larger than
$R_d$. Their sample yields a significantly shorter average $\torb = 0.76$ Gyr than
what we find, probably due to systematic differences in
how \rmax\ is determined. If so, then we scale their results to estimate
\begin{equation}
\rmax \sim (3.6\pm 0.6) \frac{1}{0.76}R_d = (4.7\pm0.8)R_d.\label{e:kkg_scale}
\end{equation}
Following eq.~\ref{e:lambda}, we have $\lambda = 0.034$ for $\torb = 1$ Gyr.
Since this scales from an estimate that avoids bulges, it produces a longer
$R_d$ scale length and hence higher $\lambda$ value than assuming all the light
comes from an exponential disc.

In comparison, measurements of typical average spin parameters of halos created in cosmological simulations range from $\lambda = 0.03$ to 0.055 with each simulation set producing a broad distribution that is close to log normal and consistently having a width $\sigma_{\log(\lambda)} \approx 0.21$ to 0.23 dex found \citep{cl96,bullock+01,bett+07,maccio+07}. Our first estimate, $\lambda = 0.020$ (assuming pure exponential discs) is somewhat below the expectations of cosmological simulations, while the second estimate $\lambda = 0.034$ \citep[from scaling the results of ][]{kkg02} is at the low end of the expectations from simulations. The 0.23 dex difference in these estimates is indicative of the systematic uncertainty involved. In addition, neglect of the gaseous disc, or equivalently assuming the same distribution for it as the stars, will underestimate the angular momentum of the baryons, and thus $\lambda$. Improved estimates of $\lambda$ can be made with better modelling of the baryonic mass and angular momentum distribution in galaxies \citep[e.g.][]{og14,boo17}, but would still require assumptions about the coupling with the unseen DM halo. Our approach using eq.~\ref{e:lambda} assumes a singular isothermal sphere and $j_d/m_d=1$, both of which might introduce additional systematic bias in our estimate of $\lambda$.
 
Despite the likely systematic offset between our observational estimate of $\lambda$ (via eq.~\ref{e:lambda}) and its true value, we can nonetheless discuss the relation between the \textit{relative} scatter in \torb\ and $\lambda$, or, equivalently, the absolute scatter in $\log\,\torb$ and $\log\lambda$. The observed scatter in $\torb(\rmax)$ of $\sim$0.16 dex has several sources: (1) the physical dispersion in $\lambda$, (2) measurement uncertainties in \rmax\ and \vmax, (3) variations in the ratio $\rmax/R_d$, (4) variations in $j_d/m_d$, (5) deviations from the assumed iso-thermal density profile\footnote{this includes variations in the ratio of circular velocity measured over the disc to that at the virial radius, $V/V_{200}$}. We assume that (1) is the dominant source, but expect that the other sources make a non-negligible addition to the scatter of \torb. In \S\ref{s:samp} we (crudely) estimated (2) the scatter in \torb\ due to measurement errors as 0.06 dex.  Removing this (in quadrature), but neglecting (3), (4) and (5), then the scatter in $\lambda$ is $\sigma_{\log\lambda}\sim0.15\rm~dex$.  This is somewhat lower than predicted by CDM models \citep[0.23 dex][]{mdv08}. This is remarkable, given that we haven't even accounted for the scatter of \torb(\rmax) coming from the sources (3), (4) and (5).

The explanation for the relatively low scatter in \torb(\rmax) is likely two-fold. First, to the extent that the disc surface mass density at \rmax\ is similar in all galaxies\footnote{In \S\ref{ss:props}, below, we show that the surface brightness limits at \rmax\ varies greatly, but this does not preclude the corresponding mass densities just interior to where this limit is found to be similar.}, then high spin systems are likely to have their disc truncated at smaller radii relative to $R_d$ than low-spin systems. Therefore, the scatter in $\rmax/R_d$ (source 3) is negatively correlated to that of $\lambda$ (source 1), hence \textit{reducing} the scatter in $t_{orb}(\rmax)$, relative to the scatter in $t_{orb}(R_d)$, which would be similar to the scatter in $\lambda$. $R_{50}$ is the closest proxy we have to $R_d$ and we do indeed find that the scatter in $t_{orb}(\rmax)$ is less than that of $t_{orb}(R_{50})$ in both the optical and UV samples (Table~\ref{t:fitp}).  The effect is more prominent in the UV sample. Secondly, our sample is likely biased towards a more narrow range of spin parameters than present in a volume-complete sample of all DM halos. The lowest-$\lambda$ halos have little angular momentum and are more likely to be bulge dominated (i.e.\ S0 and elliptical galaxies), hence they will have little \HI, and not make it in to our samples. A rationale for our {\HI}-selection producing a bias against high $\lambda$ systems is less obvious. Effectively, all {\HI}-selected galaxies are detected in the optical and UV \citep{meurer+06,wong07}; the detection limits are not biasing the samples. More speculatively, there may be a bias against high $\lambda$ systems if the baryons they contain have not been able to cool enough for \HI\ or stars to form.

\subsection{Properties at disc galaxy outskirts}\label{ss:props}

\begin{figure*}
\includegraphics[width=\columnwidth]{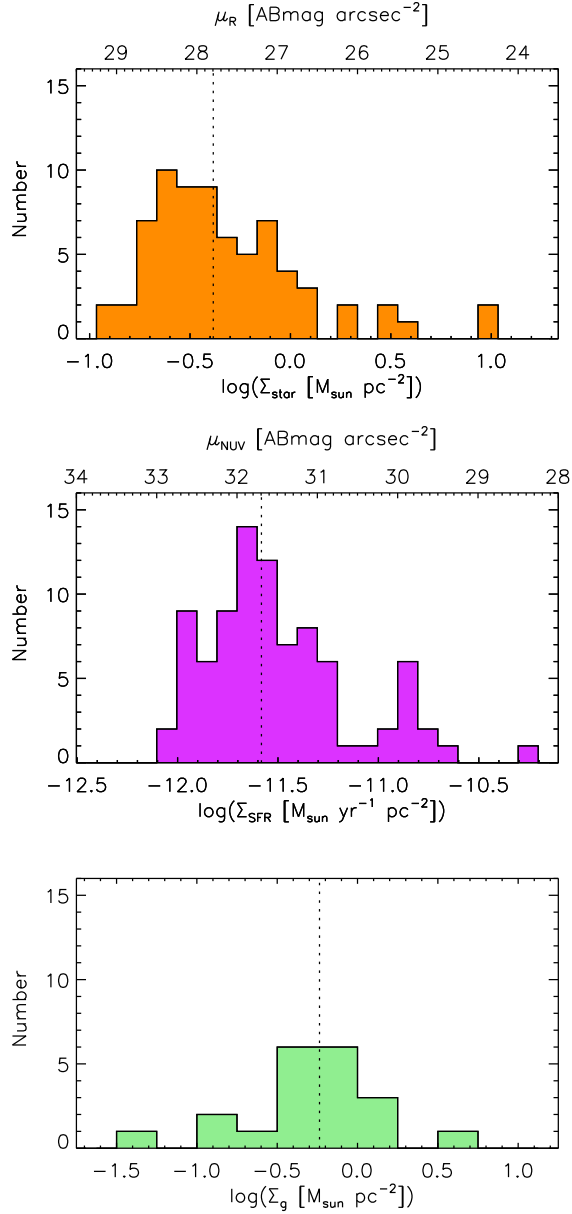}
\caption{Histograms of limiting surface brightness in the R-band and
  NUV for the SINGG and SUNGG samples are shown in the top and middle
  panels, respectively.  The bottom panel shows the $\Sigma_{\rm gas}$ at 
  the farthest point in the radial profiles of the galaxies in the \HI\ 
  sample, hence they correspond to the faintest \HI\ recorded for each galaxy. 
  The dotted vertical line in each panel shows the median of the
  distribution.  The top axis on the top and middle panels give the
  limiting surface brightness in observed units.  The bottom axes are
  calibrated in physically meaningful quantities: stellar mass density
  ($\Sigma_\star$), star formation intensity ($\Sigma_{SFR}$), and
  \HI\ surface mass density ($\Sigma_{\rm gas}$) for the top,
  middle, and bottom panels respectively.
  \label{f:limsb}}
\end{figure*}

In \S\ref{s:samp} we defined \rmax\ as the radius of readily detectable emission. It is largely determined by the amplitude of large scale ``sky variations'' in the R-band (optical) and NUV (ultraviolet). These variations represent how well we can flat field our data. The surface brightness of these variations provide a crude estimate of the limiting surface brightness at, or just interior to, \rmax. The situation is slightly different for \rmax(\HI) - the limiting surface brightness is the measured $\Sigma_{\rm HI}$ at the last measured point in published \HI\ profiles. Of course, a galaxy may extend beyond \rmax\ at fainter levels than the limiting surface brightness.  Histograms of limiting surface brightness are shown in Fig.~\ref{f:limsb}.

The bottom axes of Fig.~\ref{f:limsb} show the limiting surface
brightnesses converted to physically meaningful quantities. The R-band
surface brightness $\mu_{\rm R}$ is converted to the stellar mass density
$\Sigma_\star$ assuming a mass to light ratio
$M/L_{\rm R} = 2\, M_\odot/L_{\rm R,\odot}$. For standard IMF
assumptions, the adopted $M/L_{\rm R}$ is reasonable for a star forming
population, but probably will result in an underestimate for stellar mass
densities if the relevant stellar population is old \citep{bell+03}. To
convert $\mu_{\rm NUV}$ to star formation intensity we adopt the FUV
conversion factor of \citet{leroy+08} and assume an intrinsic colour
$\rm (FUV-NUV)_0 = 0$ ABmag, which is reasonable for the outer discs
of galaxies
\citep{thilker+07,gildepaz+07,zc07,boissier+08,hel10,werk+10a,gkr10,lee+2011}. 
The ISM density $\Sigma_g$ assumes that the ISM is dominated by
\HI\ and is corrected by a factor 1.3 to account for heavier elements.

The medians of the distributions are marked on Fig.~\ref{f:limsb} and correspond to $\Sigma_\star = 0.42\, {\rm M_\odot\, pc^{-2}}$, $\Sigma_{\rm SFR} = 2.8\times 10^{-12} {\rm M_\odot\, yr^{-1}\, pc^{-2}}$, and $\Sigma_{\rm g} = 0.58\, {\rm M_\odot\, pc^{-2}}$. These may be considered typical conditions at or near \rmax. The star formation intensity at \rmax\ is weak compared to the stellar and gaseous contents.  The time needed to form the observed stellar populations at the current star formation intensity is $t_{\rm build} = \Sigma_\star/\Sigma_{\rm SFR} = 150$ Gyr, while the time required to process the gas through star formation is $t_g = \Sigma_g/\Sigma_{\rm SFR} \sim 200$ Gyr. Equivalently both the Specific Star Formation Rate ($t_{\rm build}^{-1}$) and the Star Formation Efficiency ($t_g^{-1}$) are both low in outer discs. Thus at \rmax\ the current in situ star formation is too feeble to either create the stellar populations or transform the accumulated ISM into stars in a reasonable amount of time. 

For a galaxy to have the same \rmax\ in the R-band and the NUV implies
that the ${\rm NUV - R}$ colour at \rmax\ is similar to the ``colour'' of the
large scale sky fluctuations, i.e.\ ${\rm NUV - R} \approx \mu_{\rm sky}({\rm NUV}) -
\mu_{\rm sky}({\rm R}) \approx 3.8$ ABmag. This is a ``green-valley'' colour, i.e.\ intermediate between the blue and
red sequences \citep{schiminovich+07}, validating the long $t_{\rm
  build}$ we derive above. 

The slope of the $RV$ relationship in the R-band ($\beta_{\rm optical} = 1.13\pm 0.06$) is slightly steeper than in the NUV ($\beta_{\rm UV} = 1.04\pm 0.06$).  Comparison of Fig.~\ref{f:rvop} and Fig.~\ref{f:rvuv} shows that the values of \rmax\ in the optical and UV are nearly equal at the high end, where $V \gapeq 200\, {\rm km\, s^{-1}}$, while for $V \lapeq 50 {\rm km\, s^{-1}}$ we find that on average galaxies have $\rmax({\rm R}) \lapeq 0.7 \rmax({\rm NUV})$.  Hence at \rmax(NUV)\ galaxies are redder for large spirals than dwarfs.  This may be due to the relative importance of an old component in the disc or halo for spirals compared to dwarfs. It may also be a sign of ``down-sizing'' lower mass galaxies are less evolved in to stars than high mass galaxies.

\subsection{The edge of the disc}\label{ss:rmax}

Our results imply a distinct physical edge in the light distribution
corresponding to \rmax. The first line of evidence for this is the result that
the three tracers give nearly identical estimates of \torb; for a given \vmax\
they yield the same radius in the distribution of stars, star formation, and
atomic hydrogen. If discs were purely exponential, then the equality in \torb\
would be remarkably coincidental, since the different measurements of \rmax\ are
set by independent observational limits for each tracer. If the observational
limits were consistent within each band, then one could argue that \rmax\ is
effectively an isophotal radius. Previous studies
\citep{saintonge+08,hall+12,ss11} have shown that isophotal radii produce
tighter $RV$ relationships than those using an exponential scale length, perhaps
because of the difficulty in consistently measuring $R_d$ in the face of
contamination from the bulge, breaks in radial profiles, and biases in setting
the range of radii to fit with an exponential
\citep{freeman70,pt06,hall+12,zheng+15}. However, as shown by Fig.~\ref{f:limsb}
the limiting surface brightness is not consistent between galaxies, hence \rmax\
is not an isophotal radius. 

Indeed, the observed scatter in the $RV$
relationship provides a second line of evidence that we are dealing with a
truncation in the disc. The dispersion in limiting surface brightnesses shown in
Fig.~\ref{f:limsb} is 0.40, 1.05, 0.44 dex in the R-band, NUV, and \HI\
respectively. Assuming a pure exponential disc and adopting
Eq.~\ref{e:kkg_scale} for the mean scaling between the disc scale length and
\rmax\ then these dispersions should contribute 0.08, 0.27, and 0.09 dex to the
respective scatter in the $RV$ relationships, while the corresponding observed
scatters are 0.15, 0.16, and 0.18 dex. Thus the expected induced scatter, in
this scenario, is larger than the observed scatter in the NUV, while it would
make a considerable fraction ($\sim 25\%$ in quadrature) of the observed scatter
in the R-band and \HI.

That we are seeing a real edge to the disc is most apparent in the \HI\ sample. Using the data from MZD13 we find an average power law index $\gamma = -4.6\pm 0.5$ for $\Sigma_{\rm HI}(R)$ profiles between $R_2$ and \rmax(HI) (the uncertainty is the standard error on the mean).  If this slope is maintained towards larger $R$, the total \HI\ content is well constrained. This is unlike the region $R_1$ to $R_2$, where the \HI\ traces DM well but $\gamma \approx -1$, which can not be maintained indefinitely. Modern \HI\ observations are sufficiently deep that large improvements in sensitivity of observations do not result in large changes to the \HI\ content.  For example, \citet{gentile+13} present HALOGAS survey data of NGC~3198 with an \HI\ surface brightness sensitivity ten times fainter than the THINGS observations used by MZD13. Those improved observations result in an increase of 6\% in the \HI\ flux, and 21\%\ (0.08 dex) in maximum radius compared to the THINGS data.

We conclude that discs are not purely exponential all the way to \rmax, but must have a steep fall off in surface brightness near \rmax.  An edge, or steep fall-off, in surface brightness has been noted in the optical by van der Kruit and collaborators \citep{vanderkruit07,kkg02} and in \HI\ by \citet{vangorkom93}. Our results are similar to theirs (Fig.~\ref{f:rvall}) indicating that disc has nearly identical truncations at \rmax\ in stars, star formation and atomic hydrogen, and that it is this physical disc truncation that we are observing. 

Baryons clearly exist beyond \rmax\ in galaxies. For example at the rotational amplitude of our Galaxy $V = 220\, {\rm km\, s^{-1}}$, then $\torb = 1$ Gyr corresponds to $\rmax = 33\, {\rm kpc}$. The RC of the Milky Way disc can be traced out to $R \approx 20$ kpc \citep{sho09,bc13,bck14}, while halo blue horizontal branch stars can be detected out to $R \approx 60$ kpc \citep{xue+08} and globular clusters out to $R \approx 100$ kpc \citep[e.g.\ Pal~3;][]{kcm09}. In M31 the stellar disc can be traced to at least $R = 40$ kpc as shown by \citet{ibata+05}.  Using their adopted RC \citep{kzs02} $\torb = 1.04$ Gyr at this radius, nicely consistent with our average \torb\ at \rmax. \citet{ibata+05} point out that additional fainter disc material may be detected out to 70 kpc, while \citet{ibata+14} show faint but prominent features at larger radii relate to the halo, which extends to at least 150 kpc, about half the virial radius of $R_{\rm vir} \approx 290$ kpc \citep{kzs02}. Clearly there are stars well beyond where \torb\ is 1 Gyr in both the Milky Way and M31.  But they are primarily located in their host's halo, rather than disc.

\subsection{Alternative mechanisms to truncate discs}\label{ss:alttrunc}
 
The cosmological approach we adopted in Sec.~\ref{ss:spin} implies that discs grow  with cosmic time (the $H_z^{-1}$ dependence) due to accretion. Disc growth is also predicted in simple semi-analytic model extensions to cosmological $N$-body simulations, albeit with weaker growth \citep{dutton+11}.  However, other mechanisms may also be at play in setting the extent of galactic discs. These include the limitations in the angular momentum in an initial proto-galactic collapse \citep{vdk87}, truncation in star formation due to disc stabilisation \citep{kennicutt89,mk01} ionisation by the UV background \citep[UVB;][]{vangorkom93}, and spreading of the disc due to internal angular momentum transfer 
\citep{roskar+08a,roskar+08b}. 

The fact that we see the linear $RV$ expected for the cosmological accretion scenario, is a strong argument in its favour. Likewise, simple semi-analytic models of galaxy evolution that incorporate accretion can account for the redshift evolution of the RV relationship and other virial scaling relations \citep{dutton+11}. However, ``smoking-gun'' observations of intergalactic gas being ``caught in the act'' of accreting on to galaxies have been elusive. In a naive interpretation of the accretion scenario, one would expect outer discs to be largely gaseous. Instead, the equality of \torb\ in the R-band and \HI\ combined with the near equality of $\Sigma_{\rm HI}$ and $\Sigma_*$ in the outer discs implied by Fig.~\ref{f:limsb} suggests that they are well evolved in to stars (albeit typically less so than inner disc). This conclusion should be considered tentative since our methods for estimating the various \rmax\ values as well as $\Sigma_{\rm HI}$ and $\Sigma_*$ are crude.

An older scenario for producing a truncated but evolved outer disc is the
concept of a rapid initial collapse of galaxies including their discs
\citep{els62,freeman70}. \citet{vdk87} shows that an initial
uniformly rotating spherical gas cloud in a potential with a flat RC that
collapses while conserving angular momentum will produce an exponential disc
that truncates at 4.5 times the disc scale length. In practise, \citet{kkg02}
found that stellar discs truncate at $\rmax \sim 3.6 R_d$, i.e.\ somewhat
smaller. However, as argued in \S\ref{ss:props}, they are likely measuring
shorter \rmax\ values than we do. Indeed, \citet{vanderkruit07} notes that the
truncations examined by \citet{kkg02} correspond to $\mu_V \sim 26.5$ to 27.5
mag arcsec$^{-2}$, brighter than our estimates of the surface brightness at
\rmax\ (Fig.~\ref{f:limsb}). The fact that \citet{vanderkruit07} often find \HI\
beyond their optical truncation radii is consistent with them underestimating \rmax\ compared to us, since our \HI\ and optical \rmax\ values are consistent. When we scale
their results to our \torb (eq.~\ref{e:kkg_scale}) then we find that the ratio
between $R_d$ and \rmax\ is a factor $4.7\pm 0.8$, consistent with what is
expected from a monolithic early collapse.

\citet{kennicutt89} note that star formation, as traced by \HII\ regions in
spiral galaxies, typically cuts-off at a radius \rhii, beyond which few bright
\HII\ regions are detected. \citet{mk01} confirmed this result with improved
observations of more galaxies. Figure~\ref{f:rvall}b plots \rhii\ for five
galaxies from \citet{mk01} which are also in the sample of MZD13. The \rhii\ values (which correspond to $\torb = 48$ to 390 Myr) are considerably smaller than the \rmax\ values in our primary samples, but similar to the break radius $R_b$ of the PS1 sample.
\citet{czb10} find that the \Halpha\ distribution of edge-on spirals typically
has a downward break at $0.7R_{90}({\rm R})$ which they note may correspond to
the \rhii\ break. This scaling is very close to the $R_b \approx 0.8 R_{90}$
scaling we find for the PS1 sample, strengthening the notion that $R_b$ and \rhii\
are related. The fact that \citet{czb10} find \Halpha\ emission beyond their
break radius and the SUNGG UV measurements continue out to
$\sim 2.2 R_{90}({\rm R})$ demonstrates that the limits of galaxies traced by
prominent \HII\ regions does not measure the full extent of star formation in galactic discs.
 
Instead, UV emission is a better tracer of star formation in outer discs. The
existence of extended UV (XUV) discs \citep{gildepaz+05,thilker+05,thilker+07}
demonstrates that star formation can extend beyond the portion of the disc
readily observed in the optical. These outer discs can also be probed using
resolved stellar populations from the ground \citep{cuillandre+01,ibata+05} or
space \citep[e.g.][]{bruzzese+15}. The close match in the RV relationships at
\rmax\ shown in Fig.~\ref{f:rvall} implies that star formation extends to the
limits of the \HI\ disc. 

One mechanism that has been promoted for limiting the extent of galaxy discs is ionisation by the UVB posited by \citet{vangorkom93} to explain the the steep decline in $\Sigma_{\rm HI}$ profiles at large $R$ in NGC~3198 and other galaxies.  The scenario was consistent with modelling of the time \citep{maloney93}. The column densities he considered are similar to or somewhat smaller than the typical $\Sigma_{\rm HI} \sim 0.1\, {\rm \Msun\, pc^{-2}}$ we find at \rmax(\HI). If ionisation by the background is setting \rmax(\HI) then one should be able to detect emission from the ionised disc beyond \rmax.  \citet{hfq97} present evidence for finding this emission in the outer disc of NGC~253.  However, other searches for ionized disc gas beyond the \HI\ edges of galaxies have not been successful \citep[e.g.][]{madsen+01,dicaire+08,hlavacek+11b,adams+11}. Recent very deep integral field spectroscopy of the outermost disc of UGC~7321 finds very low surface brightness \Halpha, consistent with ionisation by the UVB, but this emission does not extend beyond the $\Sigma_{\rm HI} \sim 0.1\, {\rm \Msun\, pc^{-2}}$ contour \citep{fumagalli+17}. While UVB may ionize the ``skin'' of \HI\ discs, ionized gas does not extend much beyond the observable \HI\ disc which marks the true maximum extent of the cool ISM disc. 

Ro{\v s}kar \etal\ (2008a,b)\nocite{roskar+08a,roskar+08b} model the interplay
between star formation and disc dynamics in isolated spiral galaxies. Their
simulations produce star formation edges like that seen seen by
\citet{kennicutt89} and \citet{mk01}, beyond which the gaseous part of the disc
has a high \citet{toomre64} disc stability parameter $Q$ and thus produces
little in situ star formation. Instead, most of the old stars at large radii
formed at smaller radii and ``migrated'' outwards due to resonances with
transient spiral features. Such a process can explain the downward breaking
surface brightness profiles, ``U'' shaped age and colour profiles commonly seen
in spiral galaxies \citep[e.g.][]{pt06,bt12,zheng+15}. The material in discs
between \rhii\ and \rmax\ may then be a combination of weak XUV disc star
formation in the $Q$ stable portion of the disc combined with outwardly
migrating older stars. While this scenario is appealing, it is not obvious how
it would result in a linear $RV$ relation largely consistent at different
wavelengths down to the dwarf galaxy regime. Low mass galaxies are also a
concern because they do not have spiral density waves that are likely to drive
radial migrations.

\subsection{Other implications}\label{ss:imp}

There is a strong relationship between the \HI\ radius and \HI\ mass in galaxies of the form 
\begin{equation}
M_{\rm HI} \propto R_{\rm HI}^2.
\end{equation}
This was emphasised recently by \citet{wang+16} who note that it has been found for samples selected in a wide variety of ways \citep{br97,vs01,swaters+02,noordermeer+05,wang+13}.  The correlation implies that the average \HI\ surface brightness within $R_{\rm HI}$ is constant.  The scatter in this relationship is $\sim$0.06 dex, tighter than our $RV$ relationship. The $RV$ relationship at \rmax\ is peripherally related to this result.  It has long been known that a maximum  $\Sigma_{\rm HI} \sim 10\, {\rm \Msun\, pc^{-2}}$ is set by the conversion of the interstellar medium into a molecular form \citep[e.g.][]{bigiel+08}, and many galaxies reach this saturation in their central regions. The outer radius adopted by \citet{wang+16} is where $\Sigma_{\rm HI} = 1\, {\rm \Msun\, pc^{-2}}$ brighter than adopted for our \HI\ sample (MZD13). This effectively limits the range of allowed average surface brightness. Within galaxies, \HI\ has a predictable distribution giving a power-law fall-off in $\Sigma_{\rm HI}$, which is apparently set by the disc maintaining a constant stability parameter \citep{mzd13,wong+16}.  The limited dynamic range of $\Sigma_{\rm HI}$, combined with the shallow power law radial profile, results in the narrow range of average $\Sigma_{\rm HI}$.

\begin{figure*}
\includegraphics[width=\columnwidth]{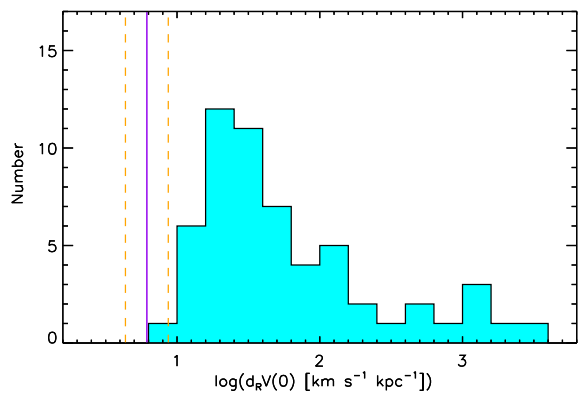}
\caption{Histogram of central velocities gradients of a wide range of disc galaxies from the sample of \citet{lfv13}. The solid vertical line marks the velocity gradient equivalent to the mean orbital time $\torb = 1$ Gyr at \rmax\ of our samples, while the dashed lines indicate the dispersion about this mean time of 0.16 dex.  Galaxies do not have central densities less than $\rho = 2.1 \times 10^{-3}\, \Msun\, {\rm pc^{-3}}$.
 \label{f:hvgrad}}
\end{figure*}

 Since radial density profiles typically decrease monotonically with $R$, the central density should not be less than the average density at \rmax.  This corresponds to a constraint on the slope of the inner RC of galaxies - the gradient should not be less than that implied by the $RV$ relation at \rmax, hence the orbital time should be less than or equal to $\sim 1$ Gyr in the central parts of galaxies.  Figure~\ref{f:hvgrad} tests this assertion by plotting the histogram of central velocity gradients, $d_RV(0)$ for 57 galaxies comprising the final sample of \citet{lfv13} with valid measurements.  The dashed line shows the gradient $d_RV(0) = 6.1\, {\rm km\, s^{-1}\, kpc^{-1}}$ corresponding to $\torb = 1\, {\rm Gyr}$.  There are no galaxies with a shallower gradient.  The shallowest $d_RV(0) = 9.0\, {\rm km\, s^{-1}\, kpc^{-1}}$ in their sample corresponds to the irregular galaxy IC~2574 \citep{deblok+08}.  Following the discussion in \S\ref{s:theory} and \S\ref{s:disc}, a galaxy with a central density less than $\langle\rho(\rmax)\rangle$ would have had to have collapsed less than our samples, and that would mean they either have a higher $\lambda$ or $j_d/m_d > 1$ (i.e.\ they have a larger fraction of the spin in the disc than the fraction of mass in the disc) or some combination of the two.  Apparently such galaxies have not (yet) formed.

\section{Conclusions}\label{s:conc}

We have shown that disc galaxies display a nearly linear radius versus velocity ($RV$) relationship at the outermost radius \rmax\ observed in the optical, ultraviolet, and \HI\ emission at 21cm. The $RV$ relationship is consistent between data sets and implies a constant orbital time of $\sim$1 Gyr at this radius. A comparison of our \HI\ selected and optically measured SINGG sample with the much larger optically selected and measured PS1 sample of \citet{zheng+15} shows nearly identical $RV$ relations at two fiducial radii.  This suggests our results are robust against the vagaries of sample selection and may be generic to disc galaxies in the low redshift Universe. Within \rmax, matter has collapsed by a factor of 37 to $\rho_M = 2.1\times 10^{-3}\, \Msun\, {\rm pc^{-2}}$, a factor $4.9\times 10^4$ times higher than the present day average matter density in the Universe.

We argue that \rmax\ in our data sets corresponds to the edge of the disc. Recent studies indicate that \rmax(\HI)\ is limited by the available ISM in the disc rather than external ionisation by the ultraviolet background. The star formation intensity at \rmax\ is an order of magnitude too weak to build up the existing stellar populations or consume the available gas within a Hubble time.  Hence, star formation at its current rate is not solely responsible for setting this radius. While \rmax\ appears to mark a sharp truncation in the disc of galaxies, it does not enclose all baryons.  Stars in the halo are distributed to much larger radii, and their kinematics indicate the dark matter also extends further, likely to the virial radius. 

Instead \rmax\ must be set by other processes such as accretion \citep[e.g.][]{sfov08,brook+12,molla+16}. Continuous cosmic accretion provides a natural explanation for the $RV$ relation. In that scenario, the $RV$ relation gives a constraint on the average spin parameter, which we estimate to be in the range $\lambda = 0.02$ to 0.035. This estimate is likely to be biased due to the crudeness of our estimates and the requirement for \HI\ in our samples which will bias them against the typically gas poor and low spin elliptical and S0 galaxies. The scatter in the orbital times provides a constraint on the dispersion of spin parameters $\sigma_{\log(\lambda)} \approx 0.16$ dex, somewhat smaller than expected by theory ($\sim 0.22$ dex), probably also due in part to the previously mentioned biases.  The scatter in orbital times may also underestimate that in $\lambda$ if \rmax\ corresponds to a consistent disc surface brightness or mass density.

An older theory, consistent with our results, is that \rmax\ is set by a rapid early collapse. Unfortunately, this scenario does not make a prediction on the $RV$ relationship. However, a crude estimate of the scaling of \rmax\ with disc scale length $R_d$ \citep[following the results of][]{kkg02} is consistent with long-standing theoretical predictions for this scenario \citep[$\rmax/R_d \sim 4.5$][]{vdk87}.  Our estimates of the gas and stellar surface densities near \rmax\ are very similar, indicating a high degree of evolution of outer discs. The relatively flat metallicity gradients in the outskirts of galaxies also indicates a high degree of chemical evolution in disc outskirts \citep{werk+10b,wpms11}. Hence an early rapid collapse model is nominally consistent with our results.  However, our estimates of $\rmax/R_d$ and the surface densities at \rmax\ are crude. Better estimates are needed to test this interpretation.  

The $RV$ relationship has some practical implications. Since the conversion of angular to physical radius is distance dependent, while the conversion of velocities is not (to first order), then one could use our $RV$ relationship to estimate distances.  However, since the observed relationship is linear it is not as powerful as the TFR where luminosity goes as orbital velocity to a power of three to four \citep[e.g.][]{meyer+08}.  Furthermore, due to the likely evolution in this relationship \citep{dutton+11} one must take care to limit its use to the local universe.

A simple $RV$ scaling relation provides a convenient tool to estimate the extent of galaxy discs. We used the \rmax\ found here in the model we developed to explain the nearly constant ratio of star formation rate (as traced in the ultraviolet) to the \HI\ mass \citep{wong+16}. Further development of this model would be useful for determining a wide range of properties along the star forming main sequence of galaxies.  Of particular relevance would be using such an approach, combined with observed column density distributions within galaxies, to model the likely cross-section of \HI\ absorbers \citep[e.g.][]{rb93,rws03,rws03err,zwaan+05,braun12}.  Similarly a realistic disc truncation radius could be usefully employed in setting the initial conditions for detailed dynamical simulations of local galaxies, or for modelling the inclusion of baryons in semi-analytic models.

\section*{Acknowledgments}

GRM acknowledges useful discussions with Tim Heckman, Brent Tully, Ken Freeman, Joss Bland-Hawthorn, Emma Ryan-Weber, and Martin Zwaan.  GRM also thanks the Center for Astrophysical Sciences of the Johns Hopkins University, and the National Astronomical Observatories of the Chinese Academy of Sciences for their hospitality during visits while this paper was being developed. ZZ is supported by the National Natural Science Foundation of China Grant No.\ 11703036. Partial funding for the SINGG and SUNGG surveys came from NASA grants NAG5-13083 (LTSA program), GALEX GI04-0105-0009 (NASA GALEX Guest Investigator grant), and NNX09AF85G (GALEX archival grant) to G.R.\ Meurer. The SINGG observations were made possible by a generous allocation of time from the Survey Program of the National Optical Astronomy Observatory (NOAO), which is operated by the Association of Universities for Research in Astronomy (AURA), Inc., under a cooperative agreement with the National Science Foundation. GALEX is a NASA Small Explorer, launched in 2003 April. We gratefully acknowledge NASA’s support for construction, operation and science analysis for the GALEX mission, developed in cooperation with the Centre National d’Etudes Spatiales of France and the Korean Ministry of Science and Technology. This research has made use of the NASA/IPAC Extragalactic Database (NED), which is operated by the Jet Propulsion Laboratory, California Institute of Technology, under contract with the National Aeronautics and Space Administration.

\bibliographystyle{mn2e}
\bibliography{mn-jour,torb}

\label{lastpage}

\end{document}